\documentclass[10pt,letterpaper,twocolumn,english]{revtex4}
\usepackage[T1]{fontenc}
\usepackage[latin9]{inputenc}
\usepackage{graphicx}

\makeatletter

\providecommand{\tabularnewline}{\\}

\usepackage{babel}
\makeatother

\begin{document}

\preprint{Draft, May 22, 2009}

\title{Shirley Interpolation - Optimal basis sets for detailed Brillouin
zone integrations revisited}

\author{David Prendergast}

\email{dgprendergast@lbl.gov}

\affiliation{The Molecular Foundry, Lawrence Berkeley National Laboratory, Berkeley,
California 94720}

\author{Steven G. Louie}

\affiliation{Department of Physics, University of California, Berkeley, California
94720}

\affiliation{The Molecular Foundry, Lawrence Berkeley National Laboratory, Berkeley,
California 94720}

\begin{abstract}
We present a new implementation of the k-space interpolation scheme
for electronic structure presented by E. L. Shirley, Phys. Rev. B
\textbf{54}, 16464 (1996). The method permits the construction of
a compact k-dependent Hamiltonian using a numerically optimal basis
derived from a coarse-grained set of effective single-particle electronic
structure calculations (based on density functional theory in this
work). We provide some generalizations of the initial approach which
reduce the number of required initial electronic structure calculations,
enabling accurate interpolation over the entire Brillouin zone based
on calculations at the zone-center only for large systems. We also
generalize the representation of non-local Hamiltonians, leading to
a more efficient implementation which permits the use of both norm-conserving
and ultrasoft pseudopotentials in the input calculations. Numerically
interpolated electronic eigenvalues with accuracy that is within 0.01
eV can be produced at very little computational cost. Furthermore,
accurate eigenfunctions - expressed in the optimal basis - provide
easy access to useful matrix elements for simulating spectroscopy
and we provide details for computing optical transition amplitudes.
The approach is also applicable to other theoretical frameworks such
as the Dyson equation for quasiparticle excitations or the Bethe-Salpeter
equation for optical responses.
\end{abstract}
\maketitle

\section{introduction}

Providing efficient access to accurate electronic structure is vital
to accelerating research in materials science and condensed matter
physics. This can be achieved directly by increasing the availablility
of computational resources, by developing faster numerical methods,
or by switching to more compact numerical representations. However,
if one adheres to existing methods and representations, one might
reasonably ask if more information can be extracted efficiently from
such approaches. In this work, we outline an efficient approach to
extracting detailed information on electronic structure for arbitrary
electron wave vector $\mathbf{k}$. This method applies not only to
periodic systems -- where $\mathbf{k}$ is well-defined -- but also
to models of aperiodic systems within the supercell approach \citep{COHEN1975}.
For example, periodic calculations are often used to simulate isolated
molecules in large supercells and disordered condensed phases are
commonly modeled using a supercell of sufficient size to contain relevant
structural features. Providing efficient access to first principles
electronic band structure and matrix elements over the entire Brillouin
zone (BZ) supports a wide range of research topics from Fermi surface
exploration in superconducting materials to detailed simulated spectroscopy
of dispersive bands or high energy excitations.

In 1996, Shirley \citep{Shirley1996PRB} outlined an approach within
effective single-particle electronic structure for constructing an
optimal basis which spans the BZ and can be used to build a compact
k-dependent Hamiltonian based on some coarse-grained reference calculations.
He applied this approach in detailed explorations of the dispersion
and spectroscopy of crystalline systems: silicon, germanium, graphite,
hexagonal boron nitride, lithium fluoride, and calcium fluoride. The
efficacy of his approach was tested by examination of electron band
structures, densities of states, dielectric properties, x-ray resonance
fluorescence and incoherent emission spectra, and photoelectron spectroscopy,
with details provided or referenced in \citep{Shirley1996PRB}. He
also used this basis in developing efficient approaches to Bethe-Salpeter
calculations for valence-band \citep{Benedict1998} and core-level
spectroscopy \citep{Shirley1998}. To our knowledge, this advantageous
approach has not been widely applied outside of Shirley's research.
In this work, we outline a generalized implementation of Shirley's
interpolation scheme, which has been incorporated as a post-processing
tool for use with the \texttt{Quantum-ESPRESSO \citep{Giannozzi2009}}
open-source electronic structure package. We hope that the outline
provided here indicates how easily Shirley interpolation might be
implemented in other electronic structure codes. Furthermore, we illustrate
that the advantages of such an approach are clear when applied to
large supercell calculations, where zone-center electronic structure
calculations are sufficient to accurately reproduce the electronic
structure throughout the BZ. This particular implementation has already
been used in simulating x-ray absorption spectra of molecules \emph{in
vacuo} \citep{Uejio2008,Schwartz2009}and in solution \citep{Schwartz2009b}.

Another commonly used interpolation scheme for electronic structure
exploits maximally localized Wannier functions\citep{Mostofi2008}.
This approach builds a set of Wannier functions to describe a band
complex and minimizes their spread in real-space. The resulting functions
can be quite localized and enable the calculation of first-principles
tight-binding parameters for use in calculations of Berry's phase
polarization\citep{Souza2000,Stengel2006}, electron transport\citep{Calzolari2004},
electron-phonon coupling\citep{Giustino2007,Giustino2007a}, and much
more. The Shirley interpolation method does not produce spatially
localized functions. However, for interpolation purposes, we show
that it is much more automatic to use and computationally less expensive
in terms of required input calculations. Also, conduction band states
may be generated as easily as valence band states. In particular,
there are no intrinsic difficulties in treating metallic systems and
no specific requirements for disentanglement of dispersive bands\citep{Souza2002}.

This paper is organized as follows: In Section \ref{sec:background}
we provide a summary of the work outlined in Ref. \citealp{Shirley1996PRB}.
In Section \ref{sec:more-efficient} and \ref{sec:non-local-potential}
we focus on the advances in our particular implementation over the
original work. Section \ref{sec:Applications} focuses on some applications
which highlight the advantages of the approach. Section \ref{sec:Comparison-with-Wannier}
differentiates Shirley interpolation from Wannier interpolation. Finally,
in Section \ref{sec:Future-Applications} we provide some potential
applications of this approach and then summarize our conclusions in
Section \ref{sec:Conclusions}.

\section{Background to the Method\label{sec:background}}

\begin{figure}
\includegraphics[clip,scale=0.5]{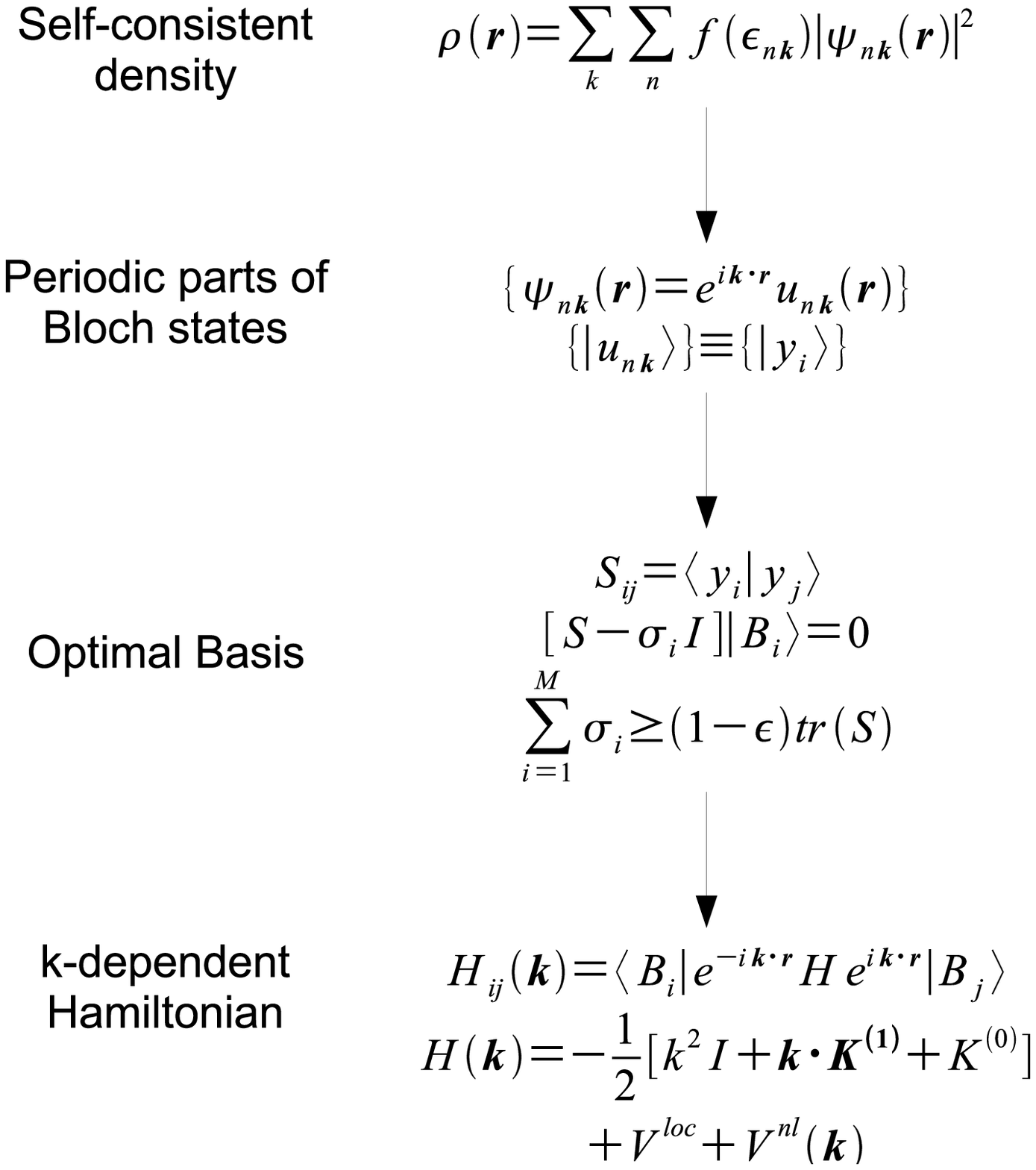}

\caption{\label{fig:flowchart}The steps involved in building the compact k-dependent
Hamiltonian in the optimal basis, beginning (top) with a self-consistent
charge density, from which states are generated for a range of bands
and k-points. An overlap matrix is constructed from the periodic parts
of these Bloch states and diagonalized. The optimal basis is chosen
as those eigenvectors of the overlap matrix (ordered by eigenvalue
magnitude) which span a user-defined fraction of the space defined
by the input states. The k-dependent Hamiltonian is constructed in
this basis from its various parts: kinetic energy, local potential,
and non-local potential.}

\end{figure}

A brief summary of Shirley's approach is provided here to establish
the context of our own work. The process of building a compact k-dependent
Hamiltonian in the optimal basis for Brillouin zone sampling is outlined
in Figure \ref{fig:flowchart}. We assume an effective single-particle
Hamiltonian (based on Kohn-Sham density functional theory\citep{Hohenberg_Kohn_1964_pr,Kohn_Sham_1965_pr}
in this work). A self-consistent charge density is generated by sufficient
sampling of the Brillouin zone. Then, if necessary, a set of states
is calculated from this density for a user-specified set of band indices
and k-points. The periodic parts of these Bloch-states are extracted
and used to construct an overlap matrix which is then diagonalized
to isolate linear dependence in this basis of periodic functions.
By ordering the overlap eigenvalues by decreasing magnitude, we may
select the optimal basis subject to a user-defined tolerance. In our
approach we truncate the basis by specifying a tolerance $\epsilon$
for the neglected fraction of the trace of the overlap.

Since we have removed the plane-wave envelope functions from the Bloch-states,
a k-dependent Hamiltonian is required, and we represent this Hamiltonian
$H(\mathbf{k})$ in the optimal basis. Each component of the Hamiltonian
is expanded as a polynomial in $\mathbf{k}$. The kinetic energy has
an analytic quadratic form, as indicated in Fig. \ref{fig:flowchart}.
Specifically, with access to the Fourier coefficients of the optimal
basis functions $B_{i}(\mathbf{G})$, if one expands the k-dependent
kinetic energy operator, one obtains:

\begin{eqnarray*}
\lefteqn{\langle B_{i}|e^{-i\mathbf{k\cdot r}}\left(-\frac{1}{2}\nabla^{2}\right)e^{i\mathbf{k\cdot r}}|B_{j}\rangle}\\
 & = & -\frac{1}{2}\left[k^{2}\delta_{ij}+\mathbf{k\cdot}\sum_{\mathbf{G}}B_{i}(\mathbf{G})^{*}\mathbf{G}B_{j}(\mathbf{G})\right.\\
 &  & \left.\hspace{3em}+\sum_{\mathbf{G}}B_{i}(\mathbf{G})^{*}G^{2}B_{j}(\mathbf{G})\right]\\
 & \equiv & -\frac{1}{2}\left[k^{2}\delta_{ij}+\mathbf{k\cdot}\mathbf{K}_{ij}^{(1)}+K_{ij}^{(0)}\right].\end{eqnarray*}

The local potential is constant with respect to $\mathbf{k}$ and
its matrix elements are efficiently computed within a plane-wave code
using Fast Fourier Transforms. The non-local potential requires fitting
on a grid in k-space. In Shirley's implementation \citep{Shirley1996PRB},
k-dependent matrix elements of the non-local potential operator were
evaluated at points on a uniform Cartesian grid which contained the
entire first Brillouin zone and these were fitted to polynomials in
$\mathbf{k}$ to enable interpolation between the points. Explicit
expressions for each term in the Hamiltonian were provided in the
original work.

The outcome of these steps is a set of coefficients which can be used
to construct the matrix $H(\mathbf{k})$ for any $\mathbf{k}$ and
then diagonalize it to produce the eigenvectors and eigenvalues in
good agreement with an equivalent solution to the underlying Hamiltonian
at the same k-point. The advantage of Shirley's approach is that one
reduces the size of the problem to be solved (i.e., the dimension
of $H$) such that detailed explorations of the eigenspectrum become
tractable. For instance, one could expect to switch from thousands
of plane-waves to perhaps tens of optimal basis functions, thereby
reducing a tough iterative (most likely parallelized) diagonalization
to a trivial direct diagonalization which can be solved efficiently
on a single processor. Shirley's satisfaction with the efficacy of
this approach is clear in his original paper and also from the large
number of detailed spectroscopic simulations enabled by it.

\section{More efficient routes to the optimal basis\label{sec:more-efficient}}

\subsection{Choice of k-points for the optimal basis}

The original scheme outlined particular uniform Cartesian grids in
k-space at which eigensolutions were generated using the DFT code
of choice and provided as input to construct the optimal basis by
diagonalization of their overlap matrix. Henceforth, we shall refer
to these DFT eigensolutions as ``input states'' and the k-space grid
on which they are calculated as the ``input grid''. In general, the
``input grid'' need not coincide with that used in the original self-consistent
DFT calculation, which generated the self-consistent charge density,
and the range of band indices for a given application may also differ,
particularly when exploring the unoccupied spectrum. Therefore, the
input states are often generated directly as solutions to the Kohn-Sham
equations for a fixed self-consistent field. In the original work,
the input grids contained the entire first Brillouin zone, with the
aim of reproducing the eigensolutions at all points in that zone.
It was not clear from the original work how the final accuracy would
be affected by particular choices of such coarse input grids. Furthermore,
the choice of input grid varied with lattice symmetry due to the differences
in the shape of the first Brillouin zone in Cartesian space. In this
work, we instead present a more general and automatic approach to
sampling k-space based on uniform grids in reciprocal lattice space,
spanning the unit cube $[0,1]^{3}$. This means that for any lattice
symmetry, the input grid of k-points may be characterized uniquely
by three integers $n_{1}\times n_{2}\times n_{3}$, much like a typical
electronic structure calculation. Furthermore, one can be sure that
this grid spans the volume of the Brillouin zone. In the original
scheme, the use of a Cartesian input grid leads to the inclusion of
some k-points lying outside the zone boundary for non-orthorhombic
cells.

\begin{figure*}
\includegraphics[scale=0.7]{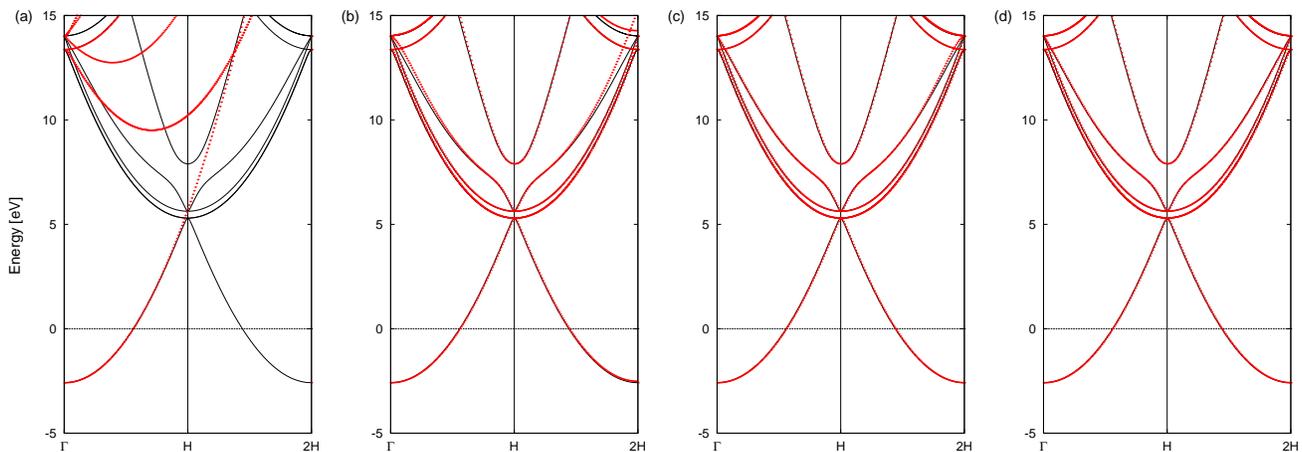}

\caption{\label{fig:sodium-bands}Accuracy in reproducing the band structure
of bcc Na with respect to choice of input k-point grid. Band energies
are reported in eV with respect to the Fermi level and k-points follow
the $\Delta$ line from the zone center ($\Gamma$) to one of the
zone boundaries ($H$) and beyond to a periodic image of the zone-center
($2H$). DFT (Shirley interpolated) band structure is indicated by
black lines (red dots). The size of the input k-point grid is varied:
(a) $\Gamma$ only; (b) $\Gamma$ and $H$; (c) $\Gamma$, $H$ and
$2H$; (d) $\Gamma$ together with its seven periodic images from
the corners of the unit cube $[0,1]^{3}$.}

\end{figure*}

\subsection{Building in periodicity with respect to $\mathbf{k}$}

As stated by Shirley, the k-dependent Hamiltonian does not impose
periodicity in k-space, and so, one must be careful regarding the
k-points one passes to the Hamiltonian for diagonalization. We illustrate
this point in Figure \ref{fig:sodium-bands} by examining the band
structure of bcc Na along the $\Delta$ line connecting the $\Gamma$
and $H$ points, with extension to the $2H$ point, which is equivalent
to $\Gamma$. We chose Na since its electronic structure is well described
at the DFT level using a local pseudopotential, thereby removing any
complications associated with interpolating the non-local potential.
In this calculation, we employed a norm-conserving pseudopotential
and a 30 Ry kinetic energy cut-off. The interpolated band structure
is clearly dependent on the choice of input k-points used to generate
the basis. Using the $\Gamma$-point alone {[}Fig. \ref{fig:sodium-bands}(a)]
results in accurate electronic bands at that point, but large errors
as one follows the $\Delta$ line. Most notably, the periodic image
of the zone-center, $2H$, is completely wrong, which emphasizes that
we have no explicit periodic boundary condition in our k-dependent
Hamiltonian. Inclusion of the zone-boundary $H$ point {[}Fig. \ref{fig:sodium-bands}(b)]
leads to marked improvement in accuracy along $\Delta$ but leads
to some inaccuracy once we leave the zone-centered first Brillouin
zone -- again the $2H$ point is not reproduced. However, the ability
to obtain excellent agreement in band structure between the input
grid points naturally prompts one to continue adding points to enable
reproducibility over a larger region of k-space. Explicitly adding
the $2H$ point {[}Fig. \ref{fig:sodium-bands}(c)] does indeed almost
restore the correct symmetry of the band structure, albeit only in
the neighborhood of this $[\Gamma,2H]$ interval -- we should expect
no reproducibility outside this interval. At this point, one can appreciate
Shirley's original Brillouin zone-spanning choice of input grids,
as they guarantee accurate reproducibility of band structure throughout
the zone. However, for cases where certain high-symmetry points are
not included in the input grid, inaccuracies may appear. 

We have observed that the input grid does not define a ``fit'' in
the usual sense of interpolation, with reproducibility decreasing
in accuracy as one explores points farther away (in the Cartesian
sense) from the grid points. In fact, when we include the $H$ point,
we notice that the entire $\Delta$ line is accurately reproduced.
Furthermore, if we were to include $\Gamma$ and its periodic image
$2H$ alone, we would see reasonable reproducibility across the entire
$\Delta$ line. Clearly, it seems that there is sufficient k-dependence
built into $H(\mathbf{k})$ to accurately describe the full-zone,
once we provide additional constraints on the symmetry via the input
states. This is equivalent to a phase-constraint for the optimal basis,
as we shall see next section. And so, in our implementation we provide
input grids chosen uniformly from the unit cube, including all corners
of the cube. We furthermore impose periodicity on k-points requested
for diagonalization by mapping them first to the unit cube (in reciprocal
lattice coordinates), since we have no guarantee of accuracy outside
the three-dimensional interval $[0,1]^{3}$.

\subsection{Accurate interpolation using just the $\Gamma$-point}

If we choose our input grid from the unit cube, it seems clear now
that we should include all corners of the cube -- $\Gamma$ and its
seven periodic images in three dimensions. Always wanting to reduce
computational effort, we immediately see that the input states originating
from these periodic images differ only in phase from those at $\Gamma$.
Furthermore, we will see that, for large supercells (small Brillouin
zones), it is sufficient, using a DFT calculation, to generate input
states at $\Gamma$ only, generating periodic images of these states
at the corners of the unit cube in a small number of steps. Figure
\ref{fig:sodium-bands}(d) illustrates how well this approach works
for bcc Na. The resulting band structure is indistinguishable (by
eye) from the original. The root-mean-square error along the indicated
path is 5.5 meV. We may reduce this error by including more input
k-points. However, we note that this error will not tend to zero,
since it is comparable to the error associated with changing the kinetic
energy cut-off and is related to a slight inconsistency in the number
of plane-waves between periodic functions from different k-points
(as noted by Shirley). In our implementation, we retain all wave vectors
\textbf{$\mathbf{G}$}, such that $\frac{1}{2}|\mathbf{k}+\mathbf{G}|^{2}\leq E_{cut}$,
padding with zeros the coefficients of those functions with missing
$\mathbf{G}$. Note that numerical differences related to this effect
reduce in magnitude for larger cut-offs, but in essence they are inconsequential,
given that the original DFT calculated eigenvalues would change as
much upon varying $E_{cut}$. Generally, for large supercells we find
that $\Gamma$-point sampling is adequate to reproduce all of the
band structure with an accuracy that is within 10 meV.

To reduce the cost of the input DFT calculation, we construct the
periodic images of the input states. The transformation of a given
periodic function $u_{n\mathbf{k}}$ to $u_{n\mathbf{k+G_{0}}}$ is
easy to obtain in plane-wave representations. Since periodicity implies
that

\begin{eqnarray*}
u_{n\mathbf{k}}(\mathbf{r})e^{i\mathbf{k.r}} & = & u_{n\mathbf{k+G_{0}}}(\mathbf{r})e^{i(\mathbf{k+G_{o}})\mathbf{.r}}\\
e^{-i\mathbf{G_{0}.r}}u_{n\mathbf{k}}(\mathbf{r}) & = & u_{n\mathbf{k+G_{0}}}(\mathbf{r})\end{eqnarray*}

then, expanding in Fourier coefficients, we have

\begin{eqnarray*}
\sum_{\mathbf{G}}c_{n\mathbf{k}}(\mathbf{G})e^{i(\mathbf{G-G_{0}})\mathbf{.r}} & = & \sum_{\mathbf{G}}c_{n\mathbf{k+G_{0}}}(\mathbf{G})e^{i\mathbf{G.r}},\end{eqnarray*}

which implies that the Fourier coefficients are ultimately reordered
according to $c_{n\mathbf{k}}(\mathbf{G+G_{0}})=c_{n\mathbf{k+G_{0}}}(\mathbf{G})$. 

For applications to large systems, where wave functions are necessarily
distributed over many processors, such reordering may be complicated
to implement. In this case, a simpler, albeit less efficient approach,
is to exploit the native implementation of the Fourier transform to
make such a transformation. Suppose that we start with the wave functions
in Fourier space. Then we follow this map

\begin{center}
$\begin{array}{ccc}
c_{n\mathbf{k}}(\mathbf{G}) & \rightarrow & u_{n\mathbf{k}}(\mathbf{r})\\
 &  & \downarrow\\
c_{n\mathbf{k+G_{0}}}(\mathbf{G}) & \leftarrow & e^{-i\mathbf{G_{0}.r}}u_{n\mathbf{k}}(\mathbf{r})\end{array}$
\par\end{center}

where we first back-transform to real-space, then multiply by the
function $e^{-i\mathbf{G_{o}.r}}$ for each $\mathbf{r}$, and then
Fourier transform again to reciprocal space.

Note that for cases where the $\Gamma$-point alone is insufficient,
this scheme could also be generalized to expand input states to the
star of a given input k-point by employing the little-group of that
k-point as determined by the lattice and atomic basis symmetry.

\section{Generalization of the non-local potential\label{sec:non-local-potential}}

\subsection{Generalized Kleinman-Bylander form}

At this point we choose an advantageous deviation from the original
implementation \citep{Shirley1996PRB}. The k-dependent non-local
potential is arbitrarily complex with respect to ${\bf k}$. Previously,
Shirley expanded the entire operator in the optimal basis on a coarse
grid in k-space and then interpolated between these values using a
polynomial expansion. This was probably the most complex component
of the original approach with respect to implementation. Details were
provided for a quartic interpolation based on a $5\times5\times5$
grid, however, the generalization to different grids and the relative
importance of such effort was left to the judgement of the reader
for specific examples. Storage of the ultimate parametrization of
the non-local potential is proportional to the grid size and the square
of the number of basis functions. In this work, we take a slightly
different approach, which we will show to be more compact in many
cases and more powerful in terms of deriving spectroscopic information.

We assume a generalized, separable, Kleinmann-Bylander \citep{KLEINMAN1982}
form for the non-local potential

\begin{eqnarray}
V^{NL} & = & \sum_{\lambda\lambda'}|\beta_{\lambda}\rangle D_{\lambda\lambda'}\langle\beta_{\lambda'}|\label{eq:}\end{eqnarray}

This is most reminiscent of Vanderbilt's ultrasoft pseudopotentials
\citep{VANDERBILT1990}. Note that for norm-conserving pseudopotentials
$D_{\lambda\lambda'}$ is diagonal. The composite index $\lambda=(I,n,l,m)$
refers to the site $I$ of a particular ion and its associated atomic
quantum numbers. Expanding the k-dependent version of this operator
in the optimal basis, we find that the k-dependence is limited to
the projectors\begin{eqnarray}
V_{ij}^{NL}(\mathbf{k\mathrm{)}} & = & \sum_{\lambda\lambda'}\beta_{\lambda i}(\mathbf{k})^{*}D_{\lambda\lambda'}\beta_{\lambda'j}(\mathbf{k}),\label{eq:}\end{eqnarray}

where

\begin{eqnarray*}
\beta_{\lambda i}(\mathbf{k}) & = & \langle\beta_{\lambda}|e^{i\mathbf{k\cdot r}}|B_{i}\rangle.\end{eqnarray*}

So, we may consider interpolating only the projector matrix elements
on a grid in k-space. The k-dependent projectors are quite efficiently
evaluated by one-dimensional Fourier transform of their radial component
and should be obtainable from the original electronic structure code.
In order to make our implementation general, we employ three-dimensional
B-spline interpolation \citep{deBoor1978,SchadowBsplines} on a uniform
$n_{1}\times n_{2}\times n_{3}$ grid of k-points in crystal coordinates,
that is, chosen from the unit cube, $[0,1]^{3}$. Requests for evaluations
of the non-local projectors at k-points outside the unit cube assume
periodicity in k-space. In general, we use a larger k-point grid to
interpolate the projectors than we use for generating the optimal
basis. A good rule of thumb is to use a grid at least twice as dense.
We note that for systems with $d$-electrons we have used more dense
grids. We also avoid spurious interpolation by limiting the order
of the B-splines to be equal to the number of grid points in each
dimension. Note that for an $n_{1}\times n_{2}\times n_{3}$ grid,
each dimension is actually expanded by one to include the edges of
the unit cell.

Note that by interpolating the projectors, one must construct the
full non-local potential for each $\mathbf{k}$ by matrix multiplication.
This apparent additional cost is not that great, considering that
for norm-conserving pseudopotentials $D_{\lambda\lambda'}$ is diagonal,
and even for ultrasoft pseudopotentials $D_{\lambda\lambda'}$ is
block-diagonal. This specific choice for interpolation can reduce
the required storage for the non-local potential by the ratio of the
number of projectors to the number of basis functions, which for many
applications is a reduction on the order of hundreds, given that there
may be typically 100 basis functions per atom, but likely less than
10 projectors.

\subsection{Extension to ultrasoft pseudopotentials}

The extension of the original approach to ultrasoft pseudopotentials
is now trivial. Ultrasoft pseudopotentials relax the norm-conservation
condition, by introducing a correction derived from atomic all-electron
and pseudo waves:

\begin{eqnarray*}
Q_{\lambda,\lambda'}(\mathbf{r}) & = & \psi_{\lambda}(\mathbf{r})^{*}\psi_{\lambda'}(\mathbf{r})-\phi_{\lambda}(\mathbf{r})^{*}\phi_{\lambda'}(\mathbf{r})\end{eqnarray*}

\begin{eqnarray*}
Q_{\lambda,\lambda'} & = & \langle\psi_{\lambda}|\psi_{\lambda'}\rangle-\langle\phi_{\lambda}|\phi_{\lambda'}\rangle\end{eqnarray*}

where $\psi,\phi$ refer to all-electron and pseudo waves respectively.
This correction appears in the coefficients that define the non-local
potential:

\begin{eqnarray*}
D_{\lambda,\lambda'} & = & D_{\lambda,\lambda'}^{ion}+D_{\lambda,\lambda'}^{Hxc},\end{eqnarray*}

where the first term is a constant for each atomic species, while
the second term involves an integral of $Q_{\lambda,\lambda'}(\mathbf{r})$
over the density dependent Hartree plus exchange-correlation potential
\citep{VANDERBILT1990}.

Furthermore, we must remember that the use of ultrasoft pseudopotentials
introduces a generalized orthonormality condition on the eigensolutions

\begin{eqnarray*}
\langle n\mathbf{k}|S|m\mathbf{k}\rangle & = & \delta_{nm},\end{eqnarray*}

which implies that their periodic components are solutions to the
following generalized eigen-problem:

\begin{eqnarray*}
[H(\mathbf{k})-\epsilon_{n\mathbf{k}}S(\mathbf{k})]|u_{n\mathbf{k}}\rangle & = & 0.\end{eqnarray*}

The k-dependent overlap matrix in the optimal basis is defined to
be

\begin{eqnarray*}
S_{ij}(\mathbf{k}) & = & \sum_{\lambda,\lambda'}\beta_{\lambda i}(\mathbf{k})^{*}Q_{\lambda,\lambda'}\beta(\mathbf{k}),\end{eqnarray*}

and the projectors matrix elements $\beta_{\lambda i}(\mathbf{k})$
are identical with those used in the non-local potential. Note that
this introduces a large storage and computational saving: we need
only interpolate the projector matrix elements once, and then we can
construct both the non-local potential and overlap matrix by multiplication.

In summary, the generalization to ultrasoft pseudopotentials has the
following additional requirements for construction of the k-dependent
Hamiltonian: (1) access to the self-consistent coefficients $D_{\lambda,\lambda'}^{Hxc}$
at the end of the SCF calculation and (2) access to the $Q_{\lambda,\lambda'}$
coefficients available in the pseudopotential definitions. Subsequently,
at any k-point, one must find the solution of a generalized eigen-problem
with both $H(\mathbf{k})$ and $S(\mathbf{k})$ constructed by interpolation.

\subsection{Advantages for spectroscopy}

One of the most common matrix elements used for optical spectroscopy
is that of the velocity operator. However, for non-local Hamiltonians,
its evaluation is non-trivial, involving a commutator of the position
operator and the non-local potential \citep{STARACE1971,KOBE1979,HYBERTSEN1987,READ1991,Kageshima1997}.
Several approaches have been introduced to include or overcome this
complication. For instance, rather than evaluating matrix elements
of velocity in the transverse gauge, one can equivalently evaluate
plane-wave matrix elements in the longitudinal gauge, exploiting the
Heisenberg equation of motion:

\begin{eqnarray*}
\langle n\mathbf{k}|\mathbf{\hat{q}\cdot v}|m\mathbf{k}\rangle & = & \lim_{q\rightarrow0}\frac{[\epsilon_{m\mathbf{k}+\mathbf{q}}-\epsilon_{n\mathbf{k}}]}{q}\langle n\mathbf{k}|e^{-i\mathbf{q\cdot r}}|m\mathbf{k}+\mathbf{q}\rangle,\end{eqnarray*}

where $\mathbf{\hat{q}}$ is the polarization of the incident electric
field. Substitution of the commutator $[e^{-iq\cdot r},H]$ in this
expression leads to the following identity:

\begin{eqnarray*}
\langle n\mathbf{k}|\mathbf{\hat{q}\cdot v}|m\mathbf{k}\rangle & = & \lim_{q\rightarrow0}\langle u_{n\mathbf{k}}|\frac{H(\mathbf{k}+\mathbf{q})-H(\mathbf{k})}{q}|u_{m\mathbf{k}+\mathbf{q}}\rangle\\
 & = & \langle u_{n\mathbf{k}}|\hat{\mathbf{q}}\cdot\frac{dH(\mathbf{k})}{d\mathbf{k}}|u_{m\mathbf{k}}\rangle.\end{eqnarray*}

One advantage of our current implementation is the ability to evaluate
derivatives of $H(\mathbf{k})$ with respect to $\mathbf{k}$. The
kinetic energy operator is readily differentiable, while the B-spline
routines we adopted also include efficient evaluation of derivatives
\citep{SchadowBsplines}. Therefore, very little extra work is required
to access optical transition amplitudes. Beyond first derivatives,
one can see that accurate effective masses are easily attainable through
the second derivatives and would be extremely efficient for large
periodic nanostructures, obviating the need for numerical differentiation
based on multiple costly k-point calculations.

\section{Applications\label{sec:Applications}}

Previous work has shown the efficacy of Shirley interpolation in reproducing
the band structure of crystalline solids. Instead, we choose to focus
on supercells, where the efficiency of Shirley's approach is clearly
competitive with the usual plane-wave calculations, while retaining
the accuracy of self-consistent results.

\subsection{Large systems using only the $\Gamma$-point}

\begin{figure}
\includegraphics[scale=0.4]{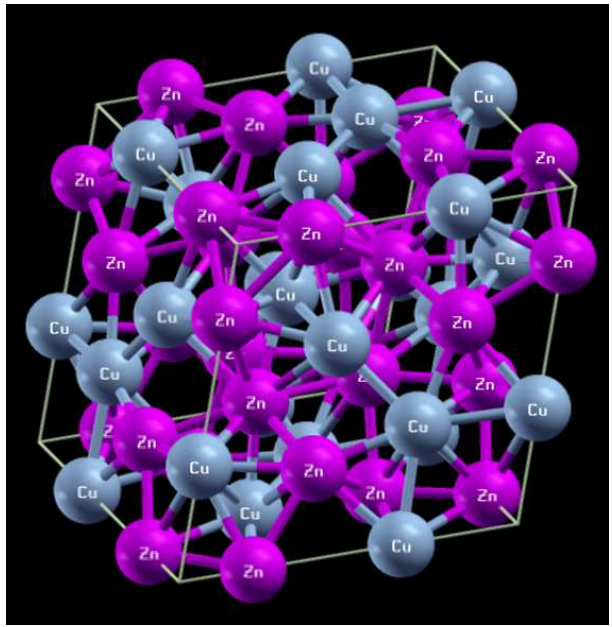}\caption{\label{fig:gamma-brass-structure}The atomic structure of $\gamma$-brass
(Cu$_{5}$Zn$_{8}$)}

\end{figure}

\begin{figure}
\includegraphics[clip,scale=0.7]{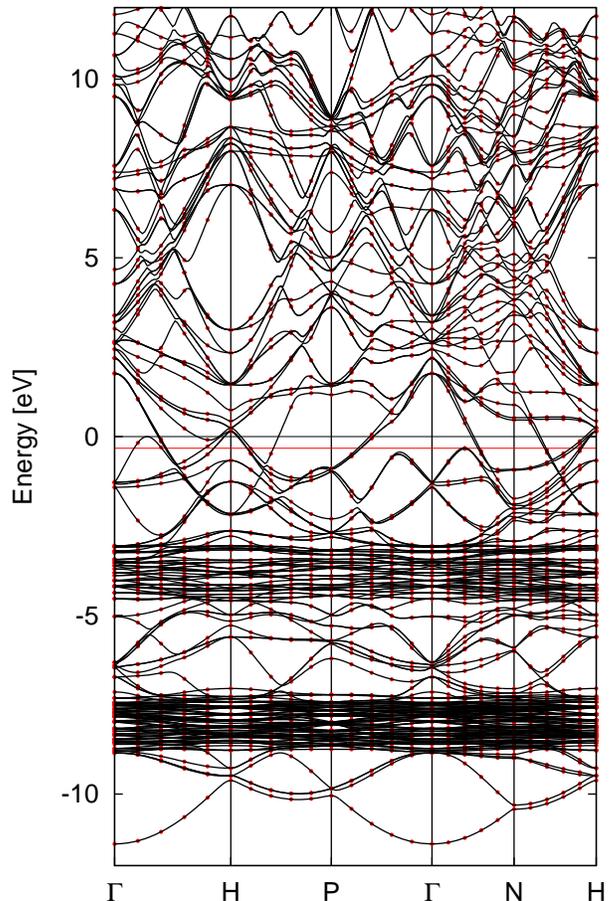}

\caption{\label{fig:gamma-brass-bands}The electron band structure of $\gamma$-brass
(Cu$_{5}$Zn$_{8}$) generated using plane-wave DFT calculations (red
dots) and using Shirley interpolation based solely on DFT wave functions
generated at the $\Gamma$-point (black lines). The root-mean-square
deviation of the Shirley interpolated values is 2 meV. The black (red)
horizontal line indicates the Fermi level resulting from a $4\times4\times4$
k-point plane-wave DFT calculation ($10\times10\times10$ k-point
Shirley interpolation).}

\end{figure}

We choose $\gamma$-brass ($\mbox{Cu}_{5}\mbox{Zn}_{8}$) as an example
of a crystalline solid with a large unit cell containing 26 atoms
(Fig. \ref{fig:gamma-brass-structure}). We generate the self-consistent
field using a shifted $4\times4\times4$ k-point grid within DFT using
ultrasoft pseudopotentials for Cu and Zn, a plane-wave cut-off of
25 Ry and a charge density cut-off of 200 Ry. The basis is built using
200 input states calculated at the $\Gamma$-point, which are then
expanded to include the seven images of the $\Gamma$-point at the
corners of the reciprocal space unit cube. The optimal basis is obtained
by diagonalizing the overlap matrix and truncating to 1095 functions
from a possible 1600, corresponding to $\sim42$ basis functions per
atom. The non-local potential is interpolated on a $3\times3\times3$
grid. The non-dispersive $d$-bands of Cu and Zn require this level
sampling in order to accurately reproduce the non-local potential.
The resulting band structure is shown in Figure \ref{fig:gamma-brass-bands}.
Comparison with DFT calculations throughout the Brillouin zone indicates
remarkable accuracy (root-mean-square deviation of 2 meV) for the
interpolated band structure, with all bands of all character ($s,p,d$)
reproduced to the same degree. We can efficiently refine a non-self-consistent
estimate of the Fermi-level using Shirley interpolation, and we find
it to be shifted from that of our initial self-consistent-field plane-wave
calculation by 0.3 eV. This is a good illustration of the importance
of detailed k-point sampling in metals. Shirley interpolation provides
an efficient route to obtaining a more accurate estimate of the Fermi
level for a given self-consistent field and indicates the possibility
of using this interpolation scheme to efficiently refine self-consistent-field
calculations for metallic systems.

\begin{figure}
\includegraphics[scale=0.5]{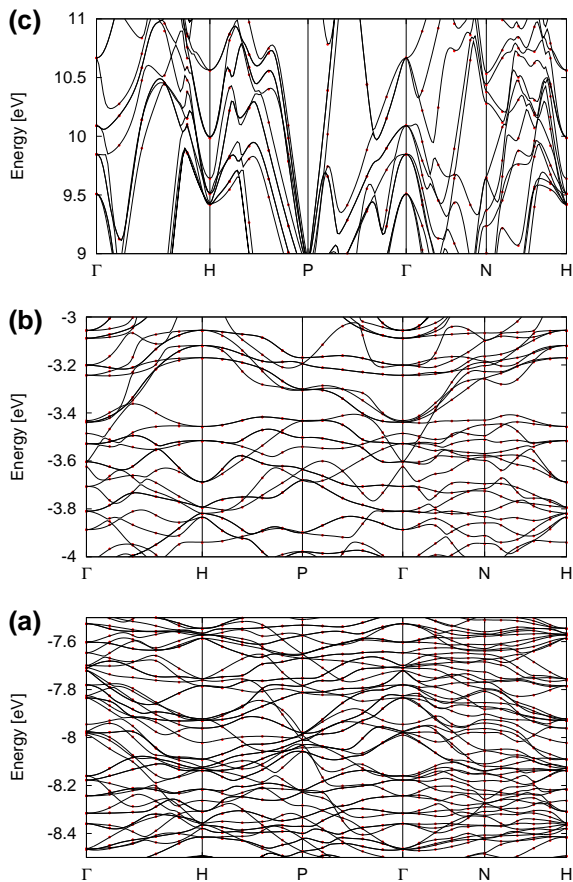}

\caption{Details of the band structure of $\gamma$-brass in three energy regions:
(a) the Zn $d$-bands; (b) the Cu $d$-bands; and (c) high in the
conduction bands. DFT plane-wave calculations are indicated by red
dots. Shirley interpolated bands are shown as black lines. Together
with Fig. \ref{fig:gamma-brass-bands} this indicates the ability
of Shirley interpolation to accurately reproduce the bands of metallic
states of varying character with minimal effort ($\Gamma$-point calculations
only).}

\end{figure}

\subsection{Efficiency \emph{in vacuo}}

When using plane-waves in supercell simulations of reduced dimensional
systems, the inclusion of large vacuum regions comes at a significant
computational cost. The use of Shirley interpolation can reduce this
cost dramatically. We use graphene as an example, where we simulate
this two-dimensional sheet of carbon atoms in a three-dimensional
supercell with a large separation between periodic images defined
by the $c$-axis. We notice that increasing $c$ results in a large
increase in the number of plane-waves in this dimension, but has only
a small impact on the number of optimal basis functions used to construct
the Shirley Hamiltonian. Table \ref{tab:graphene-timing} clearly
illustrates the efficiency of the Shirley approach for k-point sampling.
In this small example we see speed-ups of greater than 3000. Furthermore,
the increase in computational effort that we expect when adding to
the vacuum spacing is practically absent from the interpolated case
where the number of basis functions increase only slightly, rather
than the linear increase for plane-waves.

\begin{table}
\begin{tabular}{|c|c|c|c|c|}
\hline 
$c$ & $N_{PW}$ & $T_{PW}$ & $M{}_{S}$ & $T{}_{S}$\tabularnewline
\hline
\hline 
$\AA$ &  & s &  & s\tabularnewline
\hline
\hline 
10 & 21993 & 191.56 & 244 & 0.063\tabularnewline
\hline 
15 & 32971 & 220.48 & 255 & 0.056\tabularnewline
\hline 
20 & 43975 & 234.88 & 279 & 0.056\tabularnewline
\hline
\end{tabular}

\caption{\label{tab:graphene-timing}Timing information per k-point calculation
for graphene supercells of varying planar separation $c$, when using
a plane-wave DFT code $T_{PW}$ with a basis of $N_{PW}$ plane-waves,
and when using Shirley interpolation $T_{S}$, with $M_{S}$ basis
functions.}

\end{table}

\section{Comparison with Wannier Interpolation\label{sec:Comparison-with-Wannier}}

\begin{figure*}
\includegraphics[scale=0.6]{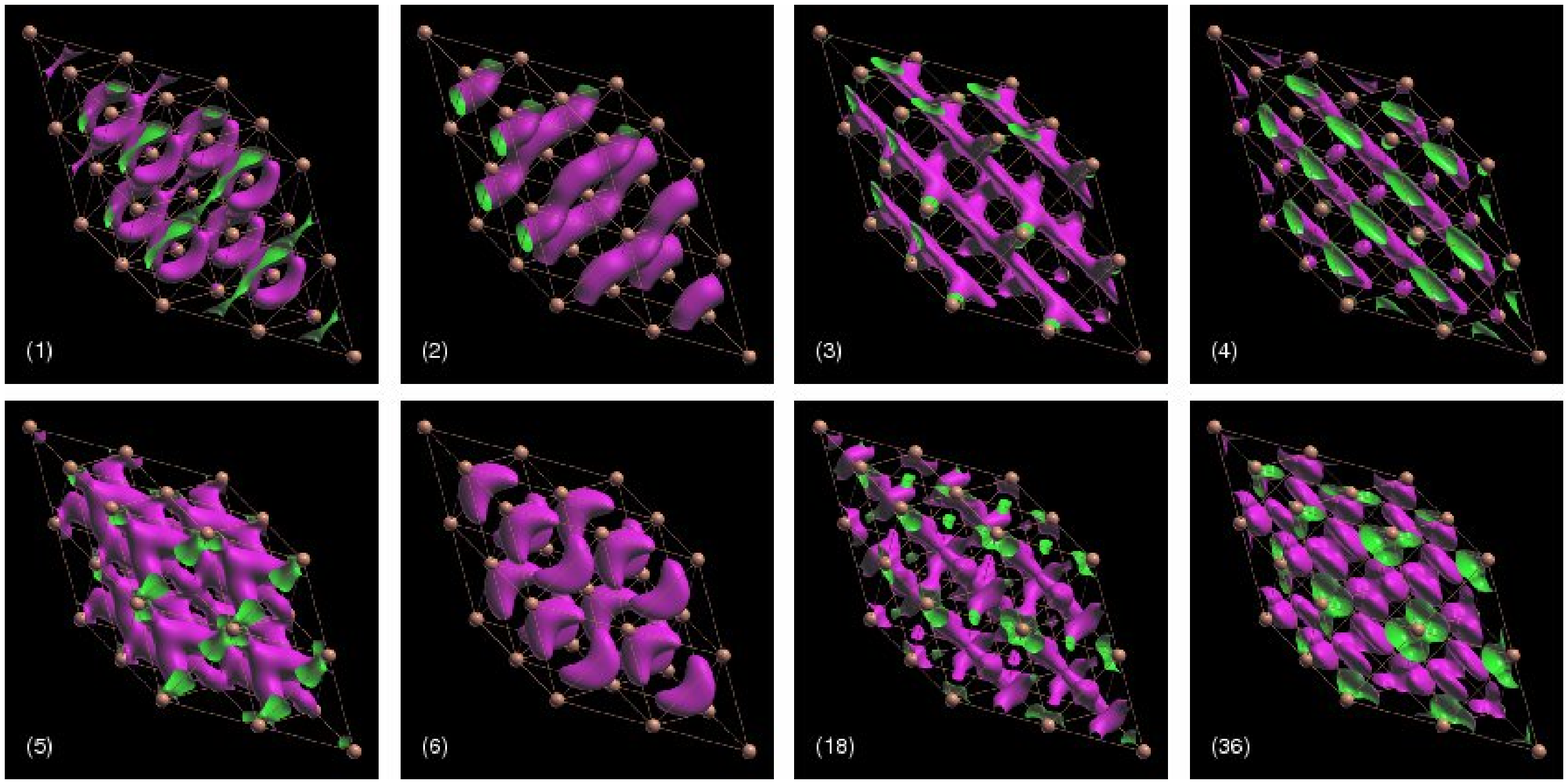}

\caption{\label{fig:Cu-optimal-basis}A subset of the optimal basis functions
of fcc Cu, determined using 8 input k-points ($\Gamma$ plus seven
images) and displayed in real-space in a $2\times2\times2$ supercell.
Copper atom positions are indicated by copper-colored spheres. Ouside
(inside) of basis function isosurfaces indicated in purple (green).
Numbers in parentheses indicate the ordering in terms of overlap matrix
eigenvalue magnitude corresponding to the following coverage of the
entire space: (1) 7.6\%; (2) 7.3\%; (3) 6.4\%; (4) 6.0\%; (5) 5.7\%;
(6) 5.4\%; (18) 1.8\%; (36) 0.78\%.}

\end{figure*}

In the last decade, the use of maximally localized Wannier functions
(MLWFs) has emerged as an extremely efficient and physically appealing
route to interpolating electronic band structure and deriving useful
tight-binding parameters from first-principles Hamiltonians \citep{Marzari1997,Mostofi2008,Giustino2007}.
The approach generates Wannier functions within a gauge which minimizes
their spatial extent. In this sense, one constructs a set of basis
functions localized in real-space, with one MLWF per band. The procedure
is similar to that outlined in Figure \ref{fig:flowchart}, with the
differences lying in an additional minimization of the spread of the
orthogonalized basis functions and no requirement to construct a k-dependent
Hamiltonian explicitly -- this is obtained rather by Fourier interpolation.
For systems with bands which do not possess an intrinsically local
character ($sp$-bands in metals, for example) the Wannierization
procedure has some intrinsic difficulties related to (i) providing
a relatively large number of k-points to enable significant localization
of the functions and (ii) disentangling such dispersive bands from
manifolds of different character which they may easily cross due to
their large dispersion. The latter problem was solved by Souza, Marzari,
and Vanderbilt \citep{Souza2002}, while the former remains an intrinsic
limitation imposed by the physical properties of the system under
study. It is particularly problematic in spectroscopic studies where
large numbers of unoccupied bands (which are in general dispersive)
are needed.

In contrast with Wannier functions, the optimal basis functions used
in Shirley interpolation have no constraint on their localization.
They are simply the result of a diagonalization of the overlap matrix
for the entire set of input periodic functions. In this sense, generating
the optimal basis functions is quite automatic and does not suffer
from issues common to most multidimensional minimization methods,
such as trapping in local minima or sensitivity to initial conditions.
The resulting functions can in fact be quite delocalized and, in general,
we have not paid much attention to their spatial dependence, given
that we do not try to exploit it in anyway. For instance, one could
not hope to extract tight-binding parameters from a set of basis functions
which are infinitely extended. Figure \ref{fig:Cu-optimal-basis}
shows a small number of the optimal basis functions derived for fcc
Cu. They are clearly delocalized, and what look like simple functions
for the larger eigenvalues of the overlap matrix become increasingly
complex for smaller eigenvalues due to the requirements of orthonormality.

Shirley interpolation is particularly suited to exploration of metallic
band structure, due to the robust automatic nature of generating the
basis and the obviation of disentangling procedures. Furthermore,
one can generate the band structure with very few initial k-points.
In fact, we have already seen that for large supercells the $\Gamma$-point
is sufficient to generate bands which accurately reproduce DFT calculations.
Wannier interpolation requires more k-points to generate accurate
band structure for bands which do not have an intrinsically localized
character. For small (monatomic) primitive cells, this may not be
problematic, but for larger supercells, where k-point sampling may
still be necessary, then there are clear advantages to using Shirley
interpolation.

Finally, it is worth mentioning that a combination of Shirley interpolation
with the Wannierization procedure may be particularly effective for
systems with intrinsic electron delocalization. Provided that one
can generate a converged self-consistent charge density, one might
use Shirley interpolation to efficiently generate solutions to the
Kohn-Sham equations at as many k-points as desired and from these
construct the necessary overlap matrix elements to begin the Wannierization
procedure. This may prove particularly advantageous for the interpolation
of metallic or high-energy unoccupied bands in large systems, such
as metallic alloys, conducting polymers, etc.

\section{Future Applications\label{sec:Future-Applications}}

The particular advantages of reducing the dimensions of a k-dependent
Hamiltonian via an optimal basis are clear for explorations of band
structure and spectroscopy. In this section, we outline further possibilities
for improved algorithms or improved scaling in both DFT and beyond-DFT
approaches.

Some self-consistent field calculations rely quite heavily on numerically
converged Brillouin zone integrations. For example, an accurate determination
of the Fermi-level is vital to an accurate estimation of the charge
density in metallic systems, and charge transfer at metallic surfaces.
For large system sizes, these calculations can prove prohibitively
expensive, since the overall cost of the calculation scales like $N_{k}N^{3}$,
where $N_{k}$ is the number of k-points and $N$ is the number of
basis functions. Even though one can assume that larger simulation
cells reduce the number of required k-points for numerical convergence,
this may not be a sufficient to reduce the overall cost significantly.
For example, doubling the system size would lead to scaling of $(N_{k}/2)(2N)^{3}=2^{2}(N_{k}N^{3})$,
which is disheartening if $N_{k}/2>1$. Since we have seen now that
for large systems one can quite easily generate accurate band energies
and states throughout the Brillouin zone, one could in principle iteratively
calculate just the zone-center electronic structure, while using interpolation
to converge the Fermi-level and self-consistent charge density upon
which the Kohn-Sham Hamiltonian is based. This would reduce the overall
scaling, removing the linear dependence on $N_{k}$ at the expense
of an increase in the overall prefactor associated with generating
the optimal basis and k-dependent Hamiltonian.

For calculating excited state properties from first principles, the
combination of the $GW$ approximation\citep{HYBERTSEN1986} and Bethe-Salpeter
Equation (BSE)\citep{Rohlfing1998,Rohlfing2000} has emerged as an
accurate and efficient approach when applied to crystalline solids,
molecules, nanostructures, and surfaces. This approach is computationally
demanding (scaling at least as $N^{4}$) and relies heavily on access
to detailed Brillouin zone sampling of the calculated electronic structure.
For periodic systems, a very fine sampling of k-space is required
to obtain converged BSE solutions, and interpolation procedures have
already been applied to enable more efficient calculations. In fact,
Shirley has already used this approach to deal with optical and x-ray
excitations in solids \citep{Benedict1998,Shirley1998}. The main
bottle-neck in such calculations comes from the need to access the
dielectric matrix at many k-points. We hope to apply the Shirley interpolation
scheme to improve the scaling of such calculations by representing
the dielectric matrix within the Shirley basis. This will be the subject
of future work.

\section{Conclusions\label{sec:Conclusions}}

We have presented a new implementation of the Shirley interpolation
method. The advances in our approach include: (1) a reduction in the
number of input electronic structure calculations required to construct
the optimal basis; (2) the ability to interpolate over the entire
Brillouin zone using just the zone-center as input for systems with
large unit cells; and (3) a generalization of the non-local potential
which reduces storage requirements, permits the use of both norm-conserving
and ultrasoft pseudopotentials. We provide applications of this method
to sodium, $\gamma$-brass, graphene, and copper which illustrate
its generality and robustness, particularly in treating metals. In
this regard, it is competitive with existing interpolation schemes
based on maximally-localized Wannier functions.

\begin{acknowledgments}
We are grateful to the following people for stimulating discussions:
Feliciano Giustino, Jeffrey B. Neaton, Robert F. Berger, and Pierre
Darancet. This work was supported by National Science Foundation grant
no. DMR04-39768 and by the Director, Office of Basic Energy Sciences,
Office of Science, U.S. Department of Energy, under Contract No. DE-AC02-05CH11231
through the LBNL Chemical Sciences Division and The Molecular Foundry.
Calculations were performed on: Franklin provided by DOE at the National
Energy Research Scientific Computing Center (NERSC); Lawrencium provided
by High Performance Computing Services, IT Division, LBNL; and Nano
at the Molecular Foundry.
\end{acknowledgments}

\end{document}